\documentclass[useAMS,usenatbib,usegraphicx,12pt]{mn2e}
\title[Black holes from low-spin material] 
{Massive black hole seeds from low angular momentum material} 
\author[Koushiappas, Bullock \& Dekel]
  {Savvas M. Koushiappas,$^1$ James S. Bullock,$^2$\thanks{Hubble fellow}
  Avishai Dekel,$^3$ \\
  $^1$ Department of Physics, The Ohio State University, 174
        W. 18th Ave, Columbus, OH 43210 USA; smkoush@mps.ohio-state.edu\\
  $^2$ Harvard Smithsonian Center for Astrophysics, 60 Garden St. Cambridge
        MA 02138 USA; jbullock@cfa.harvard.edu \\
  $^3$  Racah Institute of Physics, The Hebrew University, Jerusalem, 
        Israel; dekel@phys.huji.ac.il} 
\date{Released 2002 Xxxxx XX}

\pagerange{\pageref{firstpage}--\pageref{lastpage}} \pubyear{2003}

\begin{document}
\label{firstpage}
\maketitle


\def\lsim{\lower0.6ex\vbox{\hbox{$ \buildrel{\textstyle <}\over{\sim}\ $}}}
\def\gsim{\lower0.6ex\vbox{\hbox{$ \buildrel{\textstyle >}\over{\sim}\ $}}}

\def\equ#1{eq.~(\ref{eq:#1})}
\def\equnp#1{eq.~\ref{eq:#1}}
\def\se#1{\S\ref{sec:#1}}
\def\Fig#1{Fig.~\ref{fig:#1}}
\def\Figure#1{Figure~\ref{fig:#1}}

\def\beq{\begin{equation}}
\def\eeq{\end{equation}}
\def\be{\begin{equation}}
\def\ee{\end{equation}}
\def\prop{\propto}

\def\hMpc{h^{-1}{\rm Mpc}}
\def\Msun{\rm M_{\odot}{\ }}
\def\msun{\rm M_{\odot}{\ }}
\def\hMsun{h^{-1}M_{\odot}{\ }} 
\def\Mvir{M_{\rm v}} 
\def\Vvir{V_{\rm v}} 
\def\Vmax{V_{\rm m}} 
\def\Rvir{R_{\rm v}}  
\def\Cvir{c_{\rm v}} 
\def\Mv{M_{\rm v}} 
\def\Vv{V_{\rm v}} 
\def\Vm{V_{\rm m}} 
\def\Rv{R_{\rm v}} 
\def\Cv{c_{\rm v}} 
\def\rs{r_{\rm s}} 
\def\LCDM{$\Lambda$CDM } 
\def\Omegan{$\Omega_0$} 
\def\kms{\ {\rm km\,s^{-1}}} 
\def\hmpc{\ h^{-1}{\rm Mpc}} 
\def\hkpc{\ h^{-1}{\rm kpc}} 
\def\ihmpc{\ h~{\rm Mpc^{-1}}} 
\def\rt{\tilde{r}} 
\def\rd{r_{\rm d}} 
\def\Md{M_{\rm d}} 
\def\Sd{\Sigma_{\rm d}} 
\def\tvis{t_{\rm vis}}  
\def\rvis{r_{\rm vis}}  
\def\rtvis{\tilde{r}_{\rm vis}} 
\def\Mbh{M_{\rm bh}} 
\def\Msp{M_{\rm sp}} 
\def\Lsp{L_{\rm sp}} 
\def\tsf{t_{\rm sf}} 
\def\Sigmae{\Sigma_{\rm e}} 
\def\Sigmad{\Sigma_{\rm d}} 
\def\rtildeb{\tilde{r}_{\rm b}} 
\def\Me{M_{\rm e}} 
\def\Mspmin{M_{{\rm sp}}^{{\rm min}}} 
\def\Mbhmin{M_{{\rm bh}}^{{\rm min}}} 
\def\Mbhmax{M_{{\rm bh}}^{{\rm max}}} 
\def\zsp{z_{\rm sp}} 
\def\zbh{z_{\rm bh}} 
\def\omegasp{\omega_{\rm sp}} 
\def\omegabh{\omega_{\rm bh}} 
\def\Mmin{M_{\rm min}} 
\def\Mcrit{M_{\rm v}^{\rm crit}} 
\def\Macc{M_{\rm acc}} 
\def\bmu{b\,(\mu)} 
\def\Myr{{\rm Myr}} 
\def\tdyn{ t_{\rm dyn}} 
\def\tcool{t_{\rm cool}} 
\def\f0{f_0} 
\def\Tvir{T_{\rm vir}} 
\def\Tb{T_{\rm b}} 
\def\zvir{z_{\rm vir}} 
\def\Mc{M_{\rm c}} 
\def\tq{t_{\rm Q}} 
\def\deltac{\delta_{\rm c}} 
\def\Sone{\sigma(M)^2} 
\def\Stwo{\sigma(2M)^2} 
\def\tmm{t_{\rm mm}} 
\def\vs{v_{\rm s}} 
\def\Td{T_{\rm d}} 
\def\Ts{T_{\rm s}} 
\def\Lcrit{L_{\rm crit}} 
\def\Rcrit{{\cal{R}_{\rm crit}}} 
\def\Lmax{L_{\rm s}} 
\def\F{\tilde f} 
\def\L{\tilde \lambda} 
\def\M11{\rm{M_{11}}} 
\def\r3{\tilde{r}_{-3}} 
\def\tu{t_{\rm u}} 
\def\Mbh{M_{\rm bh}} 
\def\M9{M_9} 
\def\zion{z_{\rm ion}}
\def\zre{z_{\rm re}}
\def\zmer{z_{\rm mer}}
\def\fcrit{f_{\rm crit}}
\def\Deltavir{\Delta_{\rm vir}}
\def\OmegaMatter{\Omega_{\rm M}}
\def\OmegaLambda{\Omega_\Lambda}
\def\fcb{f_{\rm cb}}
\def\lb{\left (}
\def\rb{\right )}
\def\tildeDelta{\tilde{\Delta}}
\def\T300{{\rm T}_{300}}
\def\Ms{M_{\rm s}}
\def\fbh{f_{\rm bh}}

%
%
 
\begin{abstract} 

We present a 
cosmologically-motivated 
model in which the seeds of supermassive black holes form
out  of the  lowest angular  momentum  gas in  proto-galaxies at  high
redshift.   We  show that  under  reasonable  assumptions, this  leads
naturally  to a  correlation between  black hole  masses  and spheroid
properties,   as  observed  today.    We  assume   that  the   gas  in
early-forming, rare-peak haloes has a distribution of specific angular
momentum similar to  that derived for the dark  matter in cosmological
N-body simulations.   This distribution has a  significant low angular
momentum   tail,  which
implies   that  every
proto-galaxy should contain  gas that ends up in  a high-density disc.
In haloes  more massive  than a critical  threshold of $\sim  7 \times
10^7 \Msun$  at $z \sim  15$, the discs are  gravitationally unstable,
and  experience  an  efficient  Lin-Pringle viscosity  that  transfers
angular momentum outward and allows  mass inflow.  We assume that this
process continues until the first massive stars disrupt the disc.  The
seed black holes created in  this manner have a characteristic mass of
$\sim 10^5  \Msun$, roughly independent of the  redshift of formation.
This serves as a lower bound for black-hole masses at galactic centers
today.  The  comoving density of mass  in black hole  seeds grows with
time, tracking the continuous  production of critical mass haloes, and
saturates  when cosmic  reionization acts  to prevent  gas  cooling in
these low-mass  systems.  By  $z \sim 15$,  the comoving  mass density
becomes comparable  to that inferred from observations,  with room for
appropriate additional luminous growth during a later quasar accretion
phase.  The  hierarchical merger process  after $z \sim  15$ naturally
leads  to a  linear correlation  between black-hole  mass  and stellar
spheroid  mass, with  negligible black  hole masses  in disc-dominated
galaxies.  The formation  of massive seeds at high  redshifts, and the
relatively important role  of mergers in the buildup  of today's black
holes, are key elements in the proposed scenario. 

\end{abstract}

%
%

\begin{keywords} 
cosmology -- dark matter -- galaxies: formation - galaxies: structure.
\end{keywords}

%
%
 
\section{Introduction} 
\label{sec:intro} 

Detailed studies  of gas  and stellar kinematics  near the  centers of
present-day galaxies  have revealed  that almost all  galaxy spheroids
(ellipticals and  bulges of  spirals) host super-massive  black holes,
with masses $\Mbh  = (10^6 - 10^{9} )\Msun$, or  $\sim 10^{-4}$ of the
total  stellar mass  of  their parent  galaxies (e.g.   \citealt{KG01,
FM00,M98}).   Although most of  these black  holes are  not associated
with quasar activity today,  presumably they represent the now-dormant
counterparts  to the quasar-powering  engines known  to exist  at high
redshift \citep{FETAL01}.   Indeed, under reasonable  assumptions, the
luminosity function  of Active Galactic Nuclei (AGN)  can be explained
by modeling mass  accretion onto black holes that  have the same range
of masses as observed in  local spheroids (see e.g. \citealt{SW03} and
references  therein).   The  standard   expectation  is  that  once  a
primordial  galaxy is  populated with  a ``seed''  black hole  at some
early time, the black hole can grow via the accretion of available gas
and, in  the process,  give off light  in proportion to  its Eddington
luminosity.  

The seed population is an  important ingredient in any model that aims
to explain  AGN or  local black  holes, but the  origin and  nature of
these objects remain  unknown.  In this paper, we  provide a model for
producing rather  massive ($\sim 10^{4-5} \Msun$) seed  black holes at
early times from high density, low angular momentum gas in rare halos.
We  argue that  massive, early  forming  seeds help  explain the  high
redshift AGN population,  and, unlike lower mass seeds,  may give rise
to the observed correlations  between host galaxy properties and black
hole masses without the need for fine-tuned feedback.

Consider,  for  example  the  analysis of  \citet{BETAL03},  who  have
estimated a mass of $M_{\rm bh}  = (2-6) \times 10^9\Msun$ for a black
hole powering a $z=6.4$ quasar  found in the Sloan Digital Sky Survey.
For the \LCDM cosmology adopted below, the age of the universe at this
redshift  is  $840$  Myr.   Even   the  remnant  of  a  very  massive,
early-forming star of  $\sim 100 \Msun$ would have  to accrete mass at
the Eddington limit  for $17$ e-folding times in order  to grow to the
observed  mass.  For  a standard,  thin-disk accretion  model  with an
efficiency $\sim 0.1$, the e-folding  Salpeter time is $\sim 45$ Myrs,
so   an   intermediate-mass  seed   would   require   $770$  Myrs   of
accretion\footnote{ This  timescale is much larger  than the estimated
lifetimes of quasars at lower redshift (see \citealt{M03} for a recent
review).}   in order  to reach  to observed  mass, corresponding  to a
formation  redshift  for the  seed  of $z_f  \gsim  40$.   This is  an
extremely early redshift in  the context of \LCDM structure formation,
since  the minimum  halo mass  for ${\rm  H}_2$ cooling  at  this time
represents  a  $\ga  5  \sigma$  fluctuation.   While  this  does  not
completely rule out  the possibility that the seed  was small, it does
require an  extremely rare event.  A  $10^5 \Msun$ seed,  on the other
hand, would require $\sim 10$ e-folding times of accretion, and thus a
more  recent formation  redshift  $z_f \gsim  11$.   Of course  these
numbers are sensitive to certain assumptions, and may be influenced by
allowing  low efficiency  accretion  or super-Eddington  luminosities.
Nevertheless, it  is generally easier to understand  the population of
$z \gsim 6$  quasars if massive seeds are present  (see also Haiman \&
Loeb 2001 for a related discussion).

Perhaps  the  most  intriguing  clues  to the  nature  of  black  hole
formation and  growth are the  strong correlations between  black hole
masses and global  properties of their host spheroids  --- scales that
differ by  many orders of magnitude.   The black hole  mass is tightly
correlated with  the central velocity dispersion of  the spheroid, and
shows a significant correlation  with the spheroid luminosity or mass.
The theoretical  challenge is to connect  the galaxy-formation process
originating  on scales  of megaparsecs  with the  black-hole formation
processes on  sub-parsec scales.  In  addition, the epoch  of spheroid
formation is  likely much  later than the  time when the  first quasar
black holes begin  to shine, so an additional  challenge is to connect
these  two  components in  time,  in a  way  that  maintains, or  even
creates, the observed local correlations.

Several  scenarios have been  presented for  the formation  of massive
black hole seeds.   The general frameworks range from  those that rely
on the coalescence of massive stellar remnants that sink to the galaxy
centers \citep{QS90, HB95, L95, EETAL01} to one that models black hole
growth   via  ballistic   particle  capture   during   bulge  collapse
\citep{AGR01, AGR03}.   Other models are based  on the hydro-dynamical
evolution of  super-massive objects formed directly  out of primordial
gas \citep{HR93, LR94, EL95a,  HNR98,G01, BL03}.  Our model is related
most closely to this third type of scenario.
 
The key obstacle  in any model of black hole  formation is clearly the
centrifugal  barrier.   In  the  standard  cosmological  framework  of
hierarchical   structure  formation,   proto-haloes   acquire  angular
momentum by tidal torques  exerted by the background fluctuation field
\citep{H49,  P69, BE87,  PDH02a,  PDH02b}.  Haloes  obtain a  specific
angular momentum that  is on average $10-100$ times  smaller than that
required for centrifugal support at  the virial halo size (with a spin
parameter  of $\lambda \sim  0.01-0.1$, see  below).  This  is roughly
independent of mass  and time.  Thus, even after  cooling, most of the
gas can contract  only to radii 10-100 times  smaller than the initial
halo radius before it is halted in circular orbits.  The implied scale
(at  $z=0$) is  $\sim$kpc, many  orders of  magnitude larger  than the
corresponding Schwarzschild  radius, $\sim 10^{-5}  (M/10^9 \Msun)$pc.
The situation may get even  worse at earlier times because the typical
collapsing mass $M_\star(a)$ is smaller  in proportion to a high power
of the expansion factor $a  \equiv (1+z)^{-1}$.  This implies that the
ratio    of   the   centrifugally    supported   size    ($\propto   a
M_\star(a)^{1/3}$) to  the Schwarzschild radius  ($\propto M_\star(a)$
if  the black  hole mass  is  a constant  fraction of  the halo  mass)
increases dramatically at early times,  unless the sites of black hole
formation are  rare haloes much more massive  than $M_\star(a)$.  This
line of  argument indicates that  the seed black holes  originate from
material at  the very lowest end of  the angular-momentum distribution
within rare,  massive haloes at  high redshift.  This  material should
then perhaps lose all its  angular momentum via an effective mechanism
of  angular momentum  transfer,  which  would be  enhanced  in such  a
high-density environment.

\citet{EL95a} considered the  sites of black hole formation  to be the
very  high density discs  which form  in those  rare haloes  that have
extremely  low  total  angular  momentum.   Assuming  a  standard  CDM
scenario, they used  an analytic model to predict  the distribution of
halo spins and estimated the  abundance of haloes with low enough spin
for efficient  viscosity \citep{EL95b}. They argued that  there may be
enough low-spin haloes to explain the black hole abundance.  They then
adopted the  common view that  viscosity in a  differentially rotating
gaseous disc is the mechanism that transfers angular momentum outward,
thus  allowing  gas   to  contract  into  a  black   hole.   In  their
calculations, they considered hydrodynamic (or perhaps hydro-magnetic)
turbulence  to be the  viscous mechanism  responsible for  the angular
momentum  loss,   and  appealed  to   the  $\alpha$-disc  prescription
\citep{SS73}.

Our approach  is related, but  qualitatively different.  We  propose a
scenario where  seed black holes  form from material with  low angular
momentum in  all haloes  that are massive  enough to host  an unstable
self-gravitating disc.  The required high-density discs can only occur
in  haloes that collapse  at high  redshift, $z  \ga 10-20$,  when the
angular momentum  distribution of the (unprocessed) gas  should be the
most similar  to that of the  dark matter, and  before reionization or
some other feedback mechanism \citep[e.g.,][]{DS86, DW03} prevents the
growth  of  cold  discs.   The  crucial  point  is  that  every  halo,
regardless  of its  total  angular  momentum, is  expected  to have  a
distribution of specific angular momentum including a tail of low-spin
material.

The haloes we consider are  massive enough so that the baryons trapped
in their potential wells can cool and condense.  In the absence of any
other effect that  may alter the angular momentum  content of baryons,
the  lowest  spin   baryonic  material  in  each  system   ends  in  a
high-density central  region forming  a proto-galactic disc.   
As an illustrative fiducial model, we assume that the angular-momentum
distribution in  each halo  matches the ``universal''  distribution of
\citet{B01} (B01  hereafter). 
If  the angular  momentum is  approximately preserved  during baryonic
infall into  the disk, the  B01 angular momentum  distribution implies
that the inner disk becomes  self gravitating with the surface density
inversely  proportional to  the radius.   We show  below that  if this
behavior extends below the inner $\sim  2\% $ of the disc radius, then
enough  mass  may acquire  a  high  enough  density for  viscosity  to
efficiently redistribute the  angular momentum of the disc.
For the
final contraction of  disc material into a black  hole we evaluate the
angular-momentum loss by the effective kinematic viscosity that arises
from  gravitational  instabilities  \citep{LP87}.  This  mechanism  is
somewhat more efficient than  that from a typical $\alpha$-disc model.
This viscosity-driven  mass inflow cannot  continue indefinitely; once
the  disc  significantly fragments,  or  once  it  is heated  or  even
disrupted by feedback, the inflow stops.

The semi-quantitative scenario presented here can be regarder as an 
illustration of a general model based on the assumption that 
proto-galactic gaseous discs develop high density inner regions dominated
by low angular momentum material. Our current analysis is basically 
a feasibility study of such a model, and an attempt to recover its 
robust predictions and to constrain the model parameters by 
theoretical considerations as well as by the observations. 

The outline of  this paper is as follows.   In \S~\ref{sec:profile} we
discuss the  angular momentum  distribution in proto-galaxies  and the
implied   surface   density   profile   for  the   cold   discs.    In
\S~\ref{sec:viscosity} we investigate  the effect of kinetic viscosity
on the evolution  of the disc.  In \S~\ref{sec:mass}  we calculate the
resulting   seed   black   hole    mass   and   mass   function.    In
\S~\ref{sec:density} we  calculate the redshift evolution  of the seed
black  hole  mass  density.   In \S~\ref{sec:correlation}  we  make  a
preliminary attempt  to address the correlation between  the masses of
black  holes  and galactic  stellar  spheroids,  to  be pursued  in  a
subsequent paper.   In \S~\ref{sec:conc} we  summarize our conclusions
and  discuss our results.   When necessary,  we assume  a $\Lambda$CDM
cosmology, with a  Hubble constant $h=0.7$ ($H_0 =  100 h$ km s$^{-1}$
Mpc$^{-1}$),   universal  mass   density  $\Omega_{\rm   M}   =  0.3$,
cosmological  constant $\Omega_{\Lambda} =  0.7$, and  rms fluctuation
amplitude $\sigma_8 = 0.9$.

%
%
 
\section{Spin Distribution and Disc Profile} 
\label{sec:profile} 
 
We focus on  the proto-galaxies forming in the  redshift range $z \sim
10-30$.  In the standard picture of structure formation, galaxies form
as  a   uniform  mixture   of  dark  matter   and  baryons   in  small
over-densities  that  grow by  gravitational  instability.  Until  its
expansion  turns around into  a collapse,  each halo  acquires angular
momentum  through tidal  torques induced  by the  density fluctuations
around it.   The dissipationless collapse brings the  dark matter into
virial equilibrium in an extended spheroidal halo.  In the presence of
an  efficient  cooling  mechanism,   gas  cools  and  condenses  to  a
centrifugally supported disc.

Atomic line  cooling can  bring the gas  to $\sim 10^4$K,  and further
cooling  down  to  $\sim  300$K  requires the  presence  of  molecular
hydrogen.   The minimum  halo mass  for atomic  cooling is  $M_{\rm a}
\simeq 6  \times 10^7 \Msun  \, [(1+z)/18]^{-3/2}$, and  for molecular
hydrogen survival  it is $M_{{{\rm {H}}_2}} \simeq  5\times 10^6 \Msun
\,   [(1+z)/18]^{-7}$  in   the  redshift   range  $z   \simeq  13-30$
\citep{TETAL96, AETAL98}.  The haloes we consider below are just above
the limit  for atomic line  cooling and significantly above  the limit
for molecular hydrogen  cooling.  We therefore expect the  gas to cool
and  settle in  an  angular  momentum supported  disc  on a  timescale
comparable to the dynamical time:
\begin{equation}
\tdyn \equiv \left( {\rm G} \rho_{\rm vir} \right)^{-1/2} \simeq 77 \,
{\rm   Myr}  \,   \Delta_{178}(z)^{-1/2}   \,  \left(   \frac{1+z}{18}
\right)^{-3/2},
\end{equation}
where  we  have  assumed  our  standard  cosmological  parameters  and
$\Delta_{178}(z)$ is  the virial overdensity  in units of  $178$ times
the  background  density.\footnote{If  we  use  the  approximation  of
spherical top-hat collapse given  in \citet{BN98} we get $\Deltavir(z)
\simeq (18 \pi^2 +  82 x - 39 x^2) / \OmegaMatter(z)$,  where $x + 1 =
\OmegaMatter(z)=     \OmegaMatter(1+z)^3/[\OmegaMatter    (1+z)^3    +
\OmegaLambda]$ and  $\OmegaMatter(z)$ is the ratio of  the mean matter
density  to critical  density at  redshift $z$.   In  the $\Lambda$CDM
cosmology that we adopt, $\Deltavir(z=0) \simeq 337$ and $\Deltavir(z)
\rightarrow  178$  at  high  redshift, approaching  the  standard  CDM
($\OmegaMatter=1$) value.   }.  Hereafter,  due to the  small redshift
dependence  of  $\Deltavir$  at  the  redshifts  of  interest  we  set
$\Delta_{178}(z) = 1$.

Consider a halo of virial mass $\Mv$ and spin parameter\footnote{ This
spin parameter has been proposed by B01 as a practical modification of
the       conventional      spin      parameter,       defined      as
$\lambda=J\sqrt{|E|}/G\Mv^{5/2}$, where  $E$ is the  halo energy.  The
two definitions of $\lambda$ coincide for a singular isothermal sphere
truncated at $\Rv$.}  $\lambda  \equiv J [\sqrt{2} \Mv \Vv \Rv]^{-1}$,
where $J$  is the total halo  angular momentum.  The  virial radius is
$\Rv = 384 \, {\rm pc} \, M_7^{1/3} \, [(1+z)/18]^{-1}$ and the virial
velocity  is $\Vv  = 11\,  {\rm km}  \, {\rm  s}^{-1} \,  M_7^{1/3} \,
[(1+z)/18]^{1/2}$.  Here, $M_7 \equiv  \Mv /  10^7 \Msun$.   B01 found
that in each  halo the cumulative mass with  specific angular momentum
less than $j$ is described well by the function
\begin{equation} 
\label{eq:angprof} 
M(< j)= \mu \, \Mv \, \frac{j}{j_0 + j}, \hspace{0.3cm} \mu > 1.
\end{equation} 
This profile has an implicit maximum specific angular momentum 
$j_{\rm max} = j_0 / ( \mu - 1 )$, 
where  $j_0=\sqrt{2}  \Vv  \Rv  \lambda  /b(\mu)$,  with  $b(\mu)=-\mu
{\rm{ln}}(1 -  \mu^{-1}) - 1$.  The parameter  $\mu$ characterizes the
shape of the angular  momentum distribution.  When $\mu \rightarrow 1$
the deviation from a power law is pronounced with a larger fraction of
the mass  having low specific angular  momentum, while if  $\mu \gg 1$
the distribution resembles a pure power law.  The key property of this
distribution  is  that at  least  half  of the  halo  mass  is in  the
power-law regime,  $M(<j) \prop j$,  and the simulations  confirm that
the power-law behavior extends at least over two decades in mass, down
to $\sim 1\%$ of the total halo mass.  

The B01 profile  was  confirmed  by
\citet{CJ02} and  \citet{CARVF02} using  N-body  simulations, and  by
\citet{VDBETAL02}, \citet{CJY03}   and  \citet{SS03}  using
adiabatic  hydrodynamic simulations.  
These studies reveal that the B01 angular momentum distribution remains 
a good fit to haloes at redshifts as high as $z=3$.  In what follows
we will assume that the distribution holds at much higher redshifts,
$z \sim 15$.  
Since the $\Lambda$CDM universe is effectively Einstein-de Sitter beyond
$z \sim 1$, and the power spectrum is effectively a power law 
in the relevant mass range, one expects self-similarity between the 
haloes at $z \sim 3$ and at $z \sim 15$.
Therefore, the profile at $z \sim 15$ can be assumed to resemble the B01 
shape derived at $z=0$. 
Simple  theoretical  considerations  (B01,  \citealt{DETAL00, MDS02})
argue that this angular momentum  distribution is a natural outcome of
the   hierarchical  clustering   process.   A   qualitatively  similar
distribution is obtained when linear tidal torque theory is applied to
haloes shell by shell, or when the orbital angular momentum of merging
haloes turns  into spin  of the product  halo via tidal  stripping and
dynamical  friction.  Other justifications  follow from  more detailed
models \citep{MD02, VDBETAL01}. These analytic models provide addivtional 
support to the assumption that a similar average halo profile is valid 
at low and high redshifts. 

As mentioned above, the  B01  distribution  is a  good  fit to  simulation
results  for   angular  momentum   as  small  as   $\sim  1\%$  the
characteristic  angular  momentum  of  the halo (corresponding to 
$\sim 1 \% $ the characteristic disc radius.  As discussed
below (see Eq. 13),  our  calculation
assumes that the B01 form holds on scales below $\sim 2\%$ of the
characteristic angular momentum, so 
only a  mild extrapolation  of  the  low  angular momentum  tail
is required, and being a power law, this extrapolation seems plausible. 
Nevertheless, the angular momentum distribution of the gas may not 
exactly mirror that of the
dark matter.  In order to evaluate the qualitative effects of this possibility 
and the explore alternative extrapolations, we will also condier deviations
from the B01 distribution. 

\begin{figure} 
\label{fig:fig1}
\includegraphics[width=74mm]{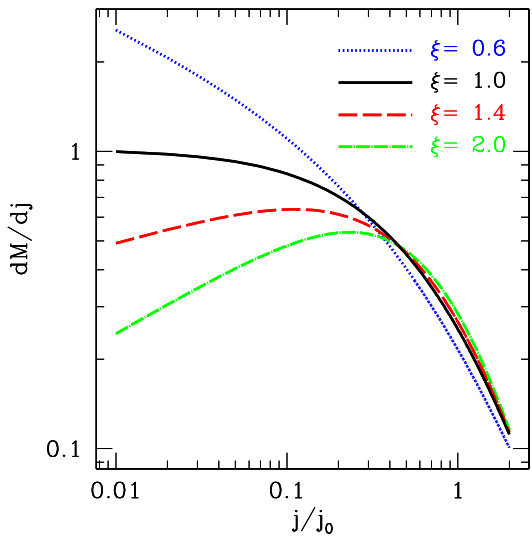}
\caption{The  mass per  unit specific  angular momentum  for different
values  of the parameter  $\xi$.  Note  that in  our fiducial  case of
$\xi=1$ (B01), this function approaches a constant at small $j$.  }
\end{figure} 

Unlike the analysis in B01 (but see \citealt{BKC02}), other works
such as \citet{CJ02}, \citet{VDBETAL02} and \citet{CJY03} focused on
estimating the fraction of dark mass with negative angular momentum
(in the direction of the total halo spin), and claimed that this fraction
may be as large as $\sim 20\%$.~\footnote{
See the extensive discussion in B01 on the pitfalls that may
arise when attempting to measure
the negative fraction of material in dark halos.}
By using the B01 profile as is we actually adopt the conservative approach
which minimizes the efficency of black-hole formation, as follows.
If a fraction of the gas starts with negative spin,
it will mix dissipatively with the positive-spin gas
leading to a purely rotating disc with no negative component.
The implications on the disc formation depends on the way by which this mixing
occurs.  In  the limiting case where the mixing occurs {\it after}
the gas has already condensed to the disc, the negative component will mix
positive spin material of exactly the same distribution of spin magnitudes.
\citep{VDBETAL02}.  This results in
a fraction of mass  with near-zero angular  momentum,
which would increase the efficiency of
black-hole production.
If angular momentum mixing occurs during the process of gas cooling and
collapse before the disc forms,
there is no  {\it a priori} reason for  the negative angular  momentum
material  to mix with  only positive angular  momentum material of the
same magnitude.  If this mixing is uniform across the angular
momentum profile, then the resulting distribution should be similar to
the  B01 form, but  with  a  slightly  smaller overall spin  parameter.
The B01 form, ignoring the negative tail, is therefore the conservative
case of minimum blak-hole formation.

We  assume  that the  specific   angular-momentum distribution  of the
baryons resembles  that of the dark matter  halo.  This assumption has
been   found   to hold  reasonably   well  in both radiative and
non-radiative hydrodynamic simulations.
\citep{VDBETAL02, CJY03}.  This assumption 
should  be valid especially  in systems
that collapse early,  as baryons would  not have been affected
by feedback associated with star formation.
In addition we assume that the angular momentum of gas is conserved as
gas cools and falls to the center forming a centrifugally supported disc. 
This assumption is backed by observations, where disc sizes seem to indicate
that the angular momentum of the gas gained by tidal torques is retained. In 
addition, B01 found that the direction of angular momentum is well aligned 
at different radii, suggesting that the formation of a disc is indeed 
possible.
Therefore, if there are no significant changes in the specific angular momentum
of the baryons during the formation of the disc,
then  the  disc angular-momentum  distribution  is
$\Md(<j)=\f0 M(<j)$,  where $\f0$ is  the final mass fraction  in cold
baryons.  For the circular orbits in  the disc one has $j(r) = rV(r)$,
so one can map the $\Md(<j)$  distribution to a radial mass profile in
the  disc.   Note  that  if  $\Md(<j)  \propto j$  (as  with  the  B01
distribution  at  small  $j$)  and  the disc  has  a  self-gravitating
rotation  curve $V^2(r)  \simeq G  \Md(r)/r$, then  $V(r)=$const.  and
$\Md(r) \propto  r$ regardless of  the initial properties of  the halo
and the possible effects of  baryonic contraction on it.  This implies
a  disc surface  density profile  of $\Sigma(r)  \prop r^{-1}$  in the
inner disc.  Since  we expect these general trends  to hold regardless
of  our  assumptions  about  the  halo profile,  we  make  the  simple
approximation  that the  final total  mass profile  of halo  plus disc
follows that of an isothermal  sphere, with $M(r) \propto r$ and $V(r)
= V_{\rm  v}$.  In  this case,  the disc mass  profile implied  by the
distribution (\ref{eq:angprof}) is
\begin{equation} 
\label{eq:massprof1} 
\Md(\rt) = f \mu \Mv \frac{\tilde{r}}{1 + \tilde{r} } \, ,
\hspace{0.5cm} \rt \equiv r/\rd \, ,
\end{equation} 
where  $\rd$ is  a characteristic  disc radius,  
\begin{equation}
\rd  \equiv 
\frac{ 
\sqrt{2}
\lambda \Rvir  }{b(\mu)} \simeq 22 {\rm pc}  \, \lambda_{0.04} 
\bmu^{-1} M_7^{1/3} \lb \frac{1+z}{18} \rb^{-1}
\end{equation}
and $\lambda_{0.04}  = \lambda / 0.04$.  At
$r \ll \rd$,  we have again $\Md \prop r$.  This  profile is valid out
to a maximum radius $\rt_{\rm{max}}  = 1/(\mu -1 )$.  The constant $f$
is the fraction of cold baryonic mass in the halo that has fallen into
the disc by redshift $z$ ($f \leq f_0$).
\begin{figure} 
\label{fig:fig2}
\includegraphics[width=74mm]{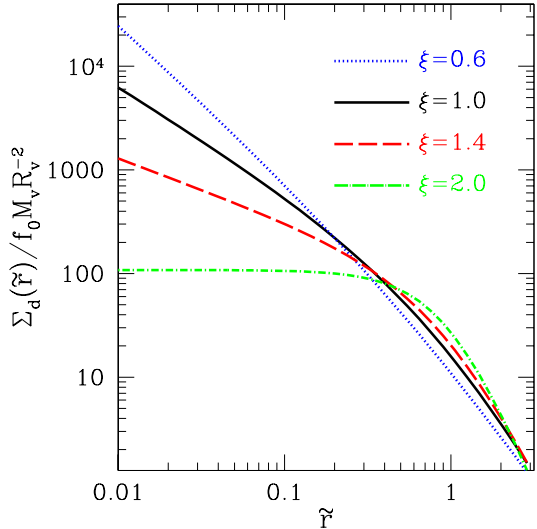}
\caption{The disc surface density profile [\equ{sigmad}] for different
values of $\xi$.   Our fiducial case, $\xi=1$, shows  a steep increase
toward small radii, while the $\xi=2$ case has a flat core, similar to
an exponential disc.  }
\end{figure} 

Recall  that it  takes a  finite time,  of order  $\sim \tdyn$,  for a
centrifugally  supported disc  to  completely form  after the  baryons
cool.  Thus  $f$ in  \equ{massprof1} becomes equal  to the  total cold
mass  fraction $\f0$ only  after a  time $t  \ge \tdyn$.   Before that
time, we assume $f= \f0 \, (t / \tdyn)$, such that the disc mass grows
linearly with time until $t = \tdyn$,
\begin{equation}
\label{eq:moft}
\Md(t) = f \Mv \simeq \f0 \Mv \lb \frac{t}{\tdyn}\rb \, , \quad 0 \leq
t \leq \tdyn \, .
\end{equation}
Hereafter we assume that the total cold mass fraction is $\f0 = 0.03$,
which is  roughly $\sim 1/5$ of  the universal baryon  fraction of the
standard   \LCDM  Universe  we   assume  in   this  paper   (see  also
\citealt{KZS02}). This is a  conservative estimate for the fraction of
baryons   that   can  form   enough   $H_2$   to  cool   significantly
\citep{TETAL96, AETAL98}.

Since our  approach relies  on an uncertain  extrapolation of  the B01
distribution, we explore a more  general form for the angular momentum
distribution, of which B01 is  a special case.  We consider the family
of  angular-momentum distributions that   correspond to  the following
disc surface density profiles, parameterized by $\xi$,
\begin{equation} 
\label{eq:sigmad} 
\Sd(\rt)=\frac{N f  \mu \Mv}{2 \pi  \rd^2} \frac{\xi \tilde{r}^{\xi
    -2} }{(1 + \rt^\xi)^2},
\hspace{0.3cm} \rt < (\mu -1 ) \, ,
\end{equation} 
with the normalization constant $N = \mu^{-1} [(\mu - 1)^\xi + 1]$.

In 
Fig.~1
we show  the differential angular  momentum distributions
for different values of $\xi$.  The  specific case of $\xi = 1$ is the
B01  form,  while  larger  (smaller)  values of  $\xi$  correspond  to
distributions  with less  (more)  low angular  momentum material.   In
Fig.~2
we show  the  resulting  disc surface  density
profiles for  different values  of $\xi$. Note  the steep  increase in
density for $\xi  = 1$ (B01), while for  $\xi=2$ the density profiles
are similar to an exponential disc.

Throughout this paper, we focus on the regime where $\tilde{r} \ll 1$,
and the surface density is well-approximated by
\begin{equation} 
\label{eq:sigmad2}
\Sd(\rt) \simeq \frac{N \xi  f \mu \Mv}{2 \pi \rd^2} \tilde{r}^{\xi
  -2}.
\end{equation} 
We  assume the  B01  limit for  the  most part  of  this paper  (which
corresponds to $\xi = 1$), but also investigate how our results change
with varying the value of $\xi$.

%
%
\section{Angular momentum loss} 
\label{sec:viscosity} 
 
After  baryons  cool and  settle  into  a  dense inner  disc,  angular
momentum  is redistributed  as a  result of  viscosity.  We  adopt the
kinematic  viscosity   prescription  proposed  by   \citet{LP87}  (see
\citealt{LR96} for a detailed test of this picture). According to this
mechanism, global gravitational instabilities in the disc give rise to
a  transfer of angular  momentum outward,  which can  be modeled  as a
local kinematic viscosity.

Consider a disc with local surface density $\Sigma$, angular frequency
$\omega$, and temperature $T$, expressed  as a sound speed $c_s \simeq
2 ({\rm  T}/300 {\rm  K})^{1/2} {\rm km  s}^{-1}$.  As  discussed by
\citet{T64}, a  disc is  gravitationally  unstable as  long as  the
minimum  size of  shear-stabilizing  disturbances $\Lmax  \simeq G  \,
\Sigma   \omega^{-2}$,   is   larger   than  the   maximum   size   of
dispersion-supported   regions,  $L_{\rm   d}   \simeq  c_s^2   G^{-1}
\Sigma^{-1}$.  This  condition is  expressed by evaluating  the Toomre
parameter $Q$,
\begin{equation}
\label{eq:Q} 
Q  \equiv \left[\frac{L_{\rm d}}{\Lmax}\right]^{1/2}  \simeq \frac{c_s
\omega}{G \Sigma}.
\end{equation}
In  discs  with  $Q<1$,  the  interaction of  the  global  instability
perturbations with  the background shear  field drives a  mass inflow.
The transfer of angular momentum occurs across regions of maximum size
$\sim \Lmax$  over a typical  timescale $\gsim \omega^{-1}$ due  to an
effective viscosity
\begin{equation}
\label{eq:nu} 
\nu    \simeq    \frac{\Lmax^2}{\omega^{-1}}    \simeq   \frac    {G^2
\Sigma^2}{\omega^3}.
\end{equation}
In what follows, we use $\omega^2 = G M(r) r^{-3}$.

A typical viscous time for mass inflow  at a radius $r$ in the disc is
$t_{\nu} \simeq r^2 \nu^{-1}$.  As  mass flows in, the central regions
of the disc  become denser, so the value of $Q$  decreases.  If at any
point $Q$ became as small as $Q^2  < H/r$, where $H \sim c_s / \omega$
is the  height of the disc,  then the instabilities  would have become
fully   dynamical,  fragmenting the   disc  into 
self-gravitating bodies, which would have stopped the angular momentum
transfer.  However, this  effect is likely to be  unimportant over the
timescales of concern  here because while  $Q$ is decreasing,
the ratio $H/r$ is actually  decreasing at a faster rate \citep{MU97}, 
thus delaying the global fragmentation and allowing the
viscosity to act  until some external process disrupts  the disc.  The
underlying assumption is that while the global instabilities give rise
to density  waves, which cause  the transfer of angular  momentum, the
disc  does not  fragment  instantaneously.  Star  formation can  occur
concurrently with  the viscous process  until feedback from  the first
massive stars  stops the gas inflow  either by heating the  disc or by
disrupting it altogether (if its virial velocity is small enough).

\subsection{Critical halo mass for viscous transport} 

Only haloes  more massive  than a critical  minimum mass  host unstable
discs in which the angular-momentum loss is efficient and a seed black
hole  can form.   This is  because, from  \equ{Q}, we  expect  $Q \sim
M^{-1/3}$,  implying a  critical mass  above  which $Q<  1$.  For  the
specific case of the B01 profile
the scaling  is $Q \propto M_d^{-1/2}  \rd^{1/2}$ $\propto M_d^{-1/3}$
$(1+z)^{-1/2}$, independent of radius.  If we assume $M_d \le \f0 \Mv$, 
then setting $\xi = 1$ in
\equ{sigmad2} yields a lower limit  for host halo masses hosting $Q<1$
discs:
\begin{eqnarray}
\label{eq:mcrit1} 
\Mcrit & \ge & 1.2 \times 10^{7} \Msun \,\T300^{3/2}
\, \lb  \frac{18}{1+z} \rb^{3/2}  \nonumber \\ &  \times &  \left[ \lb
\frac{   \lambda}{0.04  \bmu}   \rb  \lb   \frac{0.03}{f_0}   \rb  \lb
\frac{1.25}{\mu} \rb \right]^{3/2} \, .
\end{eqnarray}
All the quantities in brackets  are of order unity, and ${\rm T}_{300}
\equiv {\rm T}/300 \, {\rm K}$ is the typical temperature for gas that
has cooled via molecular Hydrogen.

A  slightly  improved estimate  of  this  critical  halo mass  can  be
obtained by considering potential  disc perturbations by major mergers
during the  gradual growth  of disc mass  until it reaches  the baryon
fraction  $\f0$  at  $t   \sim  \tdyn$  after  virialization.   If  we
approximate the rate of mass growth in the disc by \equ{moft}, then $Q
\propto   \Md^{-1/3}   \propto   t^{-1/3}$  implies   that   initially
self-gravity is negligible compared to gas pressure.  As more material
is  accreted, $Q$ decreases,  and it  may eventually  reach $Q  \le 1$
where angular momentum transfer begins.  We define the timescale $\tq$
as  the  timescale  needed  for  the disc  to  become  gravitationally
unstable had the accretion been uninterrupted.  To evaluate this time,
we assume  that the disc  is growing according to  \equ{moft}, compute
the surface density from the B01 profile (\equ{sigmad2} with $\xi=1$),
and substitute in \equ{Q} for $Q$. We then set $Q=1$ for the stability
threshold and obtain
\begin{eqnarray}
\tq & = &  1.4\, \, \tdyn \, \T300 \,  M_7^{-2/3} \, \lb \frac{18}{1+z}
\rb   \nonumber    \\   &\times&   \lb    \frac{0.03}{\f0}   \rb   \lb
\frac{1.25}{\mu} \rb \lb \frac{\lambda}{0.04 b(\mu)} \rb \, .
\end{eqnarray}

\begin{figure} 
\label{fig:fig3}
\includegraphics[width=74mm]{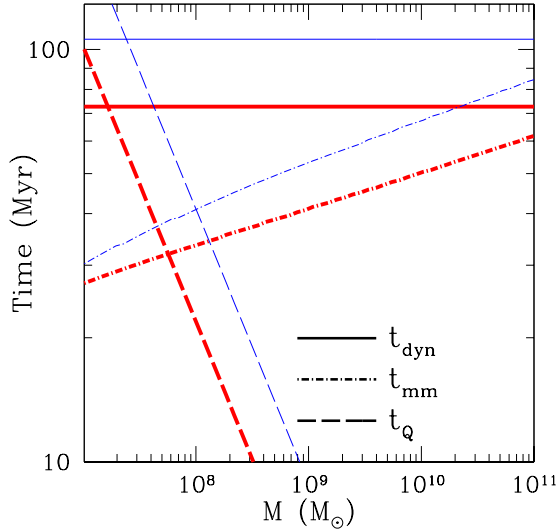}
\caption{Timescales as  a function  of halo mass.   Shown are  (a) the
time needed for  the disc to become self-gravitating  ($\tq$), (b) the
time between  virialization and  the following major  merger ($\tmm$),
and (c) the dynamical time ($\tdyn$).  Thick (red) lines correspond to
$z=17$ and thin  (blue) lines to $z=13$. The critical  mass for a halo
that can form  a seed black hole is determined  by the intersection of
the corresponding lines for $\tq$ and $\tmm$.  }
\end{figure} 

Figure 3
shows $\tq$ as a function  of host halo mass at $z=17$ and at
$z=13$. We assumed the fiducial values $\xi=1$, $T=300K$, $\f0 =0.03$,
$\mu =1.25$,  and $\lambda =0.04$.   These curves are compared  to the
virial  dynamical times  $\tdyn$ at  the two  corresponding redshifts.
The   intersection  of  the   $\tq$  and   $\tdyn$  curve   marks  the
uninterrupted  critical  mass  at   that  redshift,  as  estimated  in
\equ{mcrit1}.   However,  a major  merger  may  destroy  the disc  and
disrupt the  gas in-fall.  If the  typical time for a  major merger to
occur after the halo virializes  is $\tmm$, the disc reaches a maximum
mass of  $M_d \simeq  (\tmm/\tdyn) \f0  \Mv < \f0  \Mv$.  In  order to
obtain  a  given  critical   disc  mass  where  viscosity  efficiently
redistributes  angular momentum, we  require $\tq  < \tmm$,  i.e., the
disc should  become self-gravitating  {\it before} it  gets disrupted.
In this  case, \equ{mcrit1} is  an underestimate of the  critical halo
mass  needed for  disc  instability that  leads  to efficient  angular
momentum transfer.  In  the regime we consider, $\tmm$  is always less
than $\tdyn$.

Figure 3
also shows  the characteristic  timescale for  major mergers
$\tmm$, in the extended Press-Schechter formalism \citep{PS74, BCEK91,
LC93}.   Specifically,  we  associate  $\tmm$  with the  peak  of  the
probability distribution for  the time it would take  a halo to double
its mass  (see equation 2.21 in \citealt{LC93}).   The intersection of
the  $\tq$ and $\tmm$  curves (for  a given  redshift) marks  the more
realistic  critical minimum  mass, which  turns  out to  be roughly  a
factor   of  $\sim   4$   larger  than   the  uninterrupted   estimate
(\equnp{mcrit1}).

\begin{figure} 
\label{fig:fig4}
\includegraphics[width=74mm]{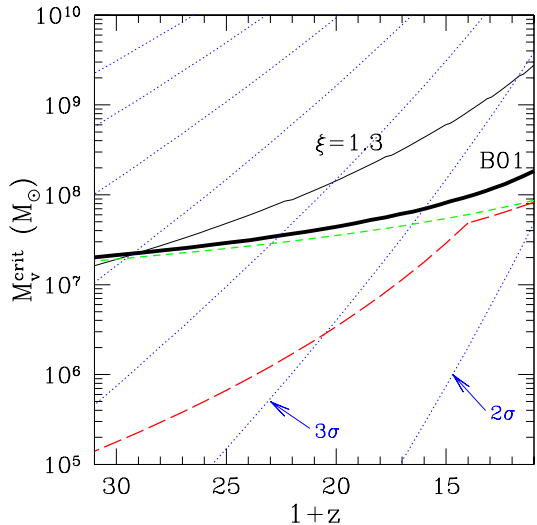}
\caption{The minimum  mass of a halo  capable of forming  a seed black
hole  as a function  of redshift.   The thick  solid curve  marked B01
refers  to our fiducial  $\xi=1$ distribution.   The thin  solid curve
refers to  $\xi=1.3$, with a  longer time of  20 Myr assumed  for star
formation.  The  dotted lines correspond to  2-8 $\sigma$ fluctuations
in the random fluctuation  field, illustrating that the minimum masses
of concern  are rare  systems.  The short  dashed line  represents the
minimum mass for atomic line  cooling only, while the long dashed line
is    an     approximation    for    molecular     hydrogen    cooling
.}
\end{figure} 

Figuer 4
shows  $M_{\rm v}^{\rm  crit}$  calculated in  the way  just
described as a  function of redshift.  The thick  solid line was 
computed assuming the B01 profile for the gas density distribution. 
 The redshift scaling is similar to that estimated in
\equ{mcrit1}, confirming that the critical mass is varying only slowly
with redshift.  The critical mass is about a factor of $\sim 4$ larger
than the uninterrupted estimate throughout the whole redshift range of
interest.  Shown for reference is  the minimum halo mass for molecular
hydrogen  to  survive  as  an  efficient  coolant;  it  justifies  our
assumption that gas can cool in critical-mass haloes.
 
Also shown  in 
Fig.~4
is  our estimated critical mass  for a flatter
surface  density profile  (or narrower  angular-momentum distribution)
with  $\xi =1.3$.   As would  be expected,  $M_{\rm v}^{\rm  crit}$ is
higher  here than  it is  for the  $\xi =  1$ case.   This  is because
densities at  small radii are lower for  a fixed mass disc,  so a more
massive host is needed in order to supply the extra disc mass required
for  $Q<1$.   Unfortunately,   the  instability  calculation  is  less
straightforward for  $\xi \ne 1$  because $Q$ is  not the same  at all
radii in the disc: $Q \propto \rt^{(1-\xi)/2}$.  Indeed, for $\xi > 1$
discs  become  unstable from  the  outside  in;  however, once  inflow
begins, mass  transport always tends  to increase the  central density
with time and therefore it is  natural to assume that the central disc
will  become unstable  within  a viscous  timescale.   We can  crudely
estimate  the critical mass  in this  case by  evaluating $Q$  and the
viscous timescale at some  characteristic radius.  If we associate the
viscous timescale  with the  timescale over which  the disc  would get
disrupted (see next section), we can define the critical halo mass for
black hole seed  formation to be the  mass scale at which $Q  < 1$ for
the given viscous timescale.  The $\xi=1.3$ curve in 
Fig.~4
has been
computed assuming  $t =  20$ Myr.  Despite  the fact that  this result
relies on very  crude approximations, it does serve  to illustrate how
narrower angular momentum distributions require more massive (and less
abundant) haloes as  the sites of black-hole formation.   We come back
to the implications of this later in the paper.

\subsection{Viscous inflow} 

Consider  now haloes  more massive  than  the critical  mass for  disc
instability  and  assume that  they  have  formed  discs with  surface
density  profiles  described  by  \equ{sigmad2}. The  viscous  process
transports  mass  inward from  a  radius  $r$  over the  viscous  time
$t_{\nu}  \simeq  r^2  \nu^{-1}$.   We  make the  assumption  that  if
viscosity has been acting for time  $t$, then all of the disc material
originally within $r_{\rm vis} = ( \nu t )^{1/2}$ has lost its angular
momentum.   Assuming  a viscosity  given  by  eq. (\ref{eq:nu}),  this
radius can be written as
\begin{eqnarray} 
\rt_{\rm  vis}^{3 - \xi} &  = &  3.9\times10^{-4} N  \, \xi^4  \, \lb
\frac{f}{0.03}   \rb   \lb   \frac{\mu}{1.25}   \rb   \lb   \frac{0.04
b(\mu)}{\lambda}  \rb^3 \nonumber  \\  & \times  & \lb  \frac{1+z}{18}
\rb^3 \lb \frac{t}{ {\rm Myr}} \rb ^2
\end{eqnarray} 
where $\rt_{\rm vis} = r_{\rm vis}/\rd$ and $f = \f0 (\tq / \tdyn)$.  
For the special case of $\xi=1$, we have
\begin{eqnarray} 
\label{eq:rvis1}
\rt_{\rm vis} & = & 1.9 \times 10^{-2} \, \lb \frac{f}{0.03} \rb^{1/2}
\lb  \frac{\mu}{1.25}  \rb   ^{1/2}  \lb  \frac{0.04  b(\mu)}{\lambda}
\rb^{3/2}  \nonumber \\ &  \times &  \lb \frac{1+z}{18}  \rb^{3/2} \lb
\frac{t}{ {\rm Myr}} \rb
\end{eqnarray}
We see that  the inflow radius grows linearly  with the time available
for viscosity to efficiently  transfer angular momentum from the inner
parts of the disc to outer radii.

According  to the  above  equations, if  the viscous  angular-momentum
transfer were to  continue uninterrupted for a long  enough time, then
the whole  disc would  have eventually lost  its angular  momentum and
collapsed.  However  as mentioned  before, the major  merger timescale
sets an  upper limit on  the time available  for this process,  and in
fact  the feedback  from star  formation  is likely  to terminate  the
process even earlier.

In our adopted  viscosity prescription, the first stars  begin to form
from  the same  instabilities  that drive  angular-momentum loss.   Of
course, if the  disc completely fragments into stars  then the angular
momentum transfer stops,  but the timescale for this  is likely longer
than the lifetimes of the  very first massive stars.  In addition, the
presence of  additional feedback on  star formation, for  example from
magnetic  fields,   may  play  a   role  in  setting   this  timescale
\citep{G01}.  Therefore,  we assume that a more  realistic estimate of
the  time  available  for  viscosity  to work  is  comparable  to  the
timescale  for collapse  and evolution  of the  first self-gravitating
bodies  in the  disc.   These  ``first stars''  should  have very  low
metallicities because  the haloes under  consideration at $z  \sim 20$
had  very few  progenitors massive  enough to  have formed  stars {\it
before } $z  \sim 20$.  Thus, the stellar evolution  in these stars is
quite rapid,  with lifetimes between $\sim  1 - 30$  Myr, depending on
many assumptions \citep{SS02, BHW01,  HW02, S02}. We explore the range
$t=  1 -  30$ Myr  for the  time available  for viscosity  to transfer
angular momentum away from the inner disc.

If angular  momentum transfer  is allowed to  operate for  a timescale
$t$, then a  large fraction of mass within  $\rt_{\rm vis}$ would lose
its angular momentum and form  a central, massive object.  This object
would be pressure-supported and short-lived
\footnote{It   is  perhaps   a  concern   that  the   presence   of  a
pressure-supported object that grows  over the viscous timescale might
influence the temperature of the  disc. We can estimate the heating of
the  disc  if  we  assume  that this  objects  radiates  at  Eddington
luminosity  with an  efficiency of  $\sim 0.1$  and that  the  disc is
heated due to Thomson scattering. We  find that it will take a time of
order $\sim 5 \tvis$ to increase the temperature of the disc by $50 \%
$ (this timescale is reduced to  $\sim 3 \tvis$ if the object radiates
at the Eddington limit).  Since  these timescales are greater than the
viscous timescales  we consider  here (set by  the timescales  for the
formation and  evolution of the  first self-gravitating bodies  in the
disc) we conclude that the presence of this central short-lived object
does not  affect the thermal  evolution of the disc.},  and inevitably
collapse  to a black  hole owing  to the  post-Newtonian gravitational
instability \citep{ST83}.  The exact  mass of the resulting black hole
depends on the mass loss experienced during the short lifetime of this
``star''.  This is an unknown quantity, but the mass loss is typically
assumed to be in the range $ 10\% - 90 \% $ of the initial mass of the
object \citep{BHW01,  HW02, SS02}.  Hereafter we assume  that the mass
of the  resulting black  hole is some  fraction $\kappa$ of  the total
mass  with zero  angular momentum,  and take  $\kappa =  0.5 $  as our
fiducial value where necessary.

%
%
 
\section{Mass function of seed black holes}
\label{sec:mass}

We define the  mass of the black hole $\Mbh$  as the fraction $\kappa$
of the  mass within $\rt_{\rm vis}(t)$,  i.e., the mass  that lost its
angular momentum over time  $t$.  Using equations (\ref{eq:rvis1}) and
(\ref{eq:sigmad2}) we obtain for the generalized profile
\begin{eqnarray} 
\label{eq:bhmass1} 
\Mbh &=& 5 \times 10^{-4} \, \Mv \, \lb \frac{\kappa}{0.5} \rb \,
\lb  \frac{ \xi  \, \,  t}{ {\rm  Myr} }  \rb^{2 \eta}  \lb
\frac{1+z}{20} \rb^{3 \eta} \\ & \times & \left[ \frac{ (N \, f \, \mu
)^{1/\xi}  \,  \bmu }{\lambda}  \right]^{3  \eta}  \left  [ 20  \right
]^{3\eta  - 3/2}  \left[ 1.1  \times  10^{-5} \right]^{2  \eta -1}  \,
,\nonumber
\end{eqnarray}   
where 
$\eta \equiv \xi / (3-\xi)$. 
For the B01 case with $\xi = 1$ the result reduces to
\begin{eqnarray} 
\label{eq:bhmass2} 
\Mbh &=& 3.8 \times 10^3  \, \Msun \lb \frac{\kappa}{0.5} \rb \lb
\frac{ f  }{0.03} \rb^{3/2} \lb  \frac{\mu \bmu}{1.25} \rb^{3/2}  \\ &
\times   &   \lb   \frac{\lambda}{0.04}   \rb^{-3/2}   \left(   \frac{
\Mv}{10^7\Msun}  \right)  \left(  \frac{1+z}{18} \right)^{3/2}  \left(
\frac{t}{  {\rm Myr}} \right) \nonumber
\end{eqnarray} 

Thus, for typical values, the black hole masses are quite large, $\sim
10^4 \Msun$,  and of order $\sim  10^{-3}$ the mass of  the host halo.
Notice  however that the  black hole  mass is  quite sensitive  to the
value  of the profile  slope parameter  $\xi$ (e.g.  the last  term in
\equnp{bhmass1}).  For example, with  $\xi = 1.4$, the black-hole mass
is $\sim  10^{-4}$ of the halo  mass.  A similar  calculation using an
exponential-like profile (with $\xi  = 2$) gives negligible seed black
hole  masses, unless  the viscous  timescale is  much larger  than the
Hubble time at that high redshift.

\begin{figure} 
\includegraphics[width=74mm]{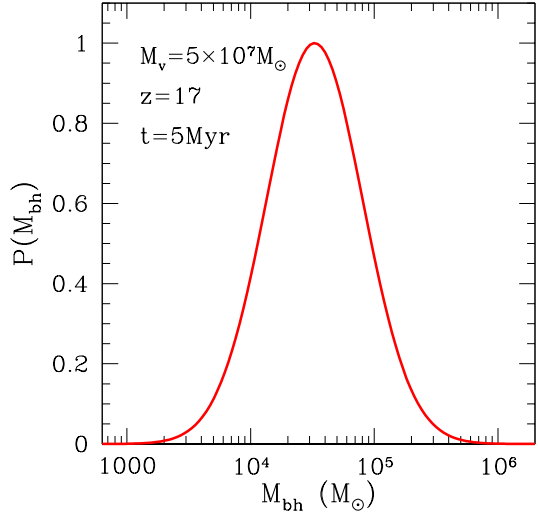}
\caption{Probability  distributions  of
seed 
black  hole  masses  for  the
threshold halo mass at $z=17$.
The assumed angular-momentum distribution is $\xi=1$, and the 
assumed viscosity timescale is 5 Myr. 
}
\label{fig:fig5}
\end{figure} 
In general, for a  halo of a given mass $\Mv$ forming  a black hole at
redshift $z$ we  expect a spread in the  parameters characterizing the
distribution  of   angular  momentum.   We   compute  the  conditional
probability distribution of seed black  hole masses given $\Mv$ at $z$
for  the $\xi =  1$ case,  where the  distributions of  the parameters
$\lambda$ and  $\mu$ are well-known.   Note first that  the black-hole
mass scales  as $\Mbh \propto [\mu  \bmu / \lambda  ]^{3/2}$, where we
define  for convenience  $\beta \equiv  \mu \bmu$  and  $\gamma \equiv
\beta/\lambda$,  such that  $\Mbh \propto  \gamma^{3/2}$.   The N-body
simulations  used by B01  show that  the probability  distributions of
$\lambda   $  and   $\beta$  are   well-parameterized   by  log-normal
distributions of the form,
\begin{equation} 
P(x) dx=\frac{1}{ \sqrt{2 \pi} \sigma_x }\frac{1}{x} {\rm{exp}} \left[
  -\frac{ {\rm{ln}}^2 (x/x_0)}{2\sigma_x^2} \right] dx
\end{equation} 
where  $x_0=0.035,  1.25$ and  $\sigma_x=0.5,  0.3$  for $x=\lambda  ,
\beta$ respectively.  Given these distributions, it is easy to compute
$P(\gamma)$, and  from this the  probability distribution of  the seed
black holes given for a halo of mass $\Mv$ at redshift $z$ for a given
viscous timescale $t$. The result is 
\begin{eqnarray} 
\label{eq:pmbh} 
P(\Mbh  |  \Mv)  \,  d\Mbh &=&  \frac{1}{\sqrt{2  \pi}  \sigma_{\Mbh}}
\frac{1}{\Mbh} \nonumber  \\ &\times& \, {\rm  exp} \left[-\frac{ {\rm
ln}^2 ( \Mbh / \Mbh^0) }{2 \sigma_{\Mbh}^2} \right] d\Mbh.
\end{eqnarray} 
where $\Mbh^0$ is
\begin{eqnarray} 
\label{eq:bhmass0} 
\Mbh^0 & = & 3.8  \times 10^4 \, \Msun \lb \frac{\kappa}{0.5} \rb
\lb \frac{  f }{0.03} \rb^{3/2} \left(  \frac{ \Mv}{10^7\Msun} \right)
\nonumber  \\ &  \times &  \left( \frac{1+z}{18}  \right)^{3/2} \left(
\frac{t}{ 10 \, {\rm Myr}} \right)
\end{eqnarray} 
and $ \sigma_{\Mbh}=0.9$.

Figure 5
shows the seed black hole 
probability  distribution normalized
to a peak  value of one, for a host halo of mass $\Mv =  5 \times 10^7
\Msun$ at $z=17$ and a time available for angular momentum transfer $t
=  5$ Myr.   In this  case, $\tdyn(z=17)  \approx 73$  Myr, 
$\tq (M=5 \times 10^5 \Msun,z=17 )
\approx 35$ Myr  and therefore $f \approx 0.014$  yielding a mean seed
black hole  mass of $\Mbh \approx  3 \times 10^4  \Msun$ with possible
masses  in  the  range  $10^3-10^6  \, \Msun$.   Such  a  distribution
function of  seed black hole  masses should provide a  useful starting
point for  studies aimed  at modeling the  quasar properties  within a
cosmological context \citep{BSF03, MHN01, KH00}  and can be used in 
of  the merger history 
of supermassive black holes \citep{MR01, VHM03}.

Consider  now black  holes forming  only  in haloes  of critical  mass
$\Mcrit$.  If we approximate $\Mcrit$ with equation (\ref{eq:mcrit1}),
then  the seed  black  hole given  in  equation (\ref{eq:bhmass1})  is
independent  of  redshift  of  formation,  host halo  mass,  and  spin
parameter, and has a ``universal'' value of
\begin{equation}
\Mbh   \simeq   6   \times   10^4   \Msun  \,   \T300^{3/2}   \,   \lb
\frac{\kappa}{0.5}  \rb \,  \lb \frac{  f_0}{0.03} \rb^{3/2}  \, \lb
\frac{t}{ 10 \, \Myr} \rb.
\end{equation}
However,  it is  expected  that  since $\Mcrit$  depends  on the  spin
parameter  $\lambda$ and  $\mu \bmu$  there will  be a  spread  in the
critical  host halo  masses.   The critical  mass  $\Mcrit$ scales  as
$\Mcrit   \propto   \gamma^{-3/2}$.   If   we   use  the   probability
distribution  for $P(\gamma)$  calculated  as explained  above, it  is
straightforward to calculate the distribution in the host halo masses.
The  result is  a  log-normal  distribution with  a  mean of  ${M^{\rm
crit,0}} \simeq  5 \times 10^7 \Msun  \, \T300^{3/2} [18/(1+z)]^{3/2}$
and  a standard  deviation of  $\sigma_{\Mcrit} =  0.9$.  Thus,  if we
focus on  black holes  forming within critical  mass haloes  only, our
model  predicts a near  universal mass  of $\sim  10^5 \Msun$,  with a
log-normal distribution of the masses hosting these seeds.

%
%

\section{Total mass in seed black holes} 
\label{sec:density}
 
In this section we calculate the comoving mass density of mass of seed
black holes and make a comparison with the observed density of mass in
black holes  today in the local  Universe.  In this  calculation we do
not include  any additional black  hole growth via  luminous accretion
(which must occur at some level  in order to match the observed quasar
population). We therefore expect  the calculated density to be smaller
than the observed value.  If, for example, most black holes experience
a luminous accretion phase characterized  by a few e-folding times, we
want our  cumulative mass density  to fall below the  locally observed
black hole density by roughly an order of magnitude.

As halo  masses grow with time,  some haloes become  more massive than
the minimum mass threshold $\Mcrit$  and become the sites of potential
seed black  hole formation.  Since  the threshold mass  decreases only
slowly with redshift (see figure \ref{fig:fig4}), the mass density due
to seed black holes is dominated by the seeds formed at the particular
redshift  of interest.  Naturally,  all haloes  more massive  than the
critical host halo  mass scale were able to form a  seed black hole at
some  higher  redshift.   However,  these mass  scales  correspond  to
higher-sigma   fluctuations   and    therefore   do   not   contribute
significantly  to the  mass density.   In  order to  include only  the
contribution coming from  seed black holes formed in  haloes that have
just crossed the minimum mass threshold, we compare the typical growth
rate of halo mass with the  growth rate of $\Mcrit$.  We find that the
probability of  a halo crossing  the threshold having  previously been
massive enough  to host black hole forming  progenitors is vanishingly
small  ($\sim 10^{-6}$).   Therefore it  is  safe to  assume that  the
contribution  to the  seed black  hole mass  density at  each redshift
comes predominantly from those seeds  formed in haloes which have just
crossed the minimum mass threshold near that particular redshift.

The comoving density of mass in  black holes as a function of redshift
can be determined by integrating  over the formation rate of haloes at
the threshold mass from some initial redshift $z_i$ to the redshift of
interest $z$, i.e.,  $\rho_{\rm bh}(z) = \int_{{z_i}}^z (d  \rho / dz)
dz$.  We approximate the evolution  in the co-moving number density of
seed black holes, as
\begin{equation}
\frac{d  \rho(z)}{dz} =  \frac{d}{dz} \left[  \int_{\Mcrit(z)} ^\infty
\frac{dn(z)}{dM} \, \Mbh(M,z) dM \right] .
\label{eq:growth}
\end{equation}
Here $[dn(z)/dM] dM$ is the number  density of haloes with mass in the
interval between $M$ and $M+dM$ at redshift $z$, given by
\begin{equation}
\frac{dn(z)}{dM}  dM  = \frac{\rho_0}{M}  \,  f_1[\sigma(M),z] \,  
\left| \frac{d \sigma^2(M)}{dM} \right| \, dM
\end{equation}
and $\sigma  (M)$ is the mean  square fluctuation amplitude  at a mass
scale $M$. Here $f_1$ is the fraction of mass associated with haloes of
mass $M$ corresponding  to the given range in  $\sigma(M)$ at redshift
$z$.   We  use  the  Press-Schechter (1974)  approximation  for  $f_1$
\citep[see,  e.g.][]{LC93}.    Since  the  value   of  $\Mcrit(z)$  is
associated  with  increasingly   higher  sigma  fluctuations  at  high
redshifts, our results  are not sensitive to the  choice of $z_i$.  We
choose $z_i = 30$, but our results are unchanged for $z_i=50$.

\begin{figure}
\label{fig:fig6}
\includegraphics[width=78mm]{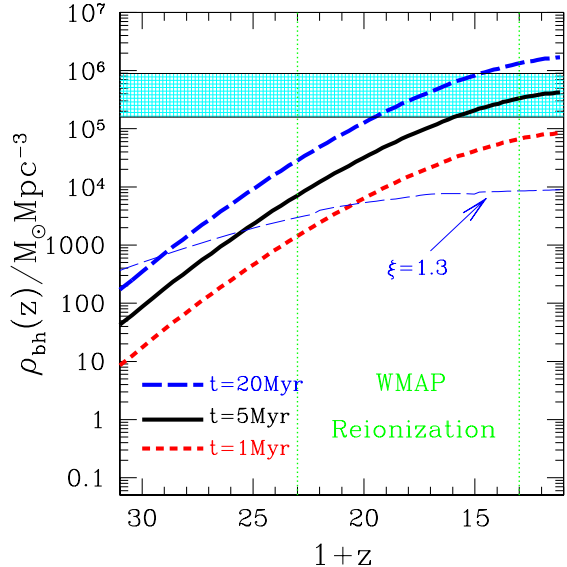}
\caption{Total  mass density  in seed  black  holes as  a function  of
redshift.    Thick  lines   represent  our   fiducial   $\xi=1$  (B01)
angular-momentum  distribution for  various viscous  timescales, while
the thin line  corresponds to $\xi=1.3$ and a  viscous timescale of 20
Myr.   The shaded  area represents  the observed  range of  total mass
density in black holes at $z=0$. The WMAP reionization range is marked
by the vertical dotted lines.}
\end{figure}
  
Figure 6
shows the comoving density  of seed black  holes obtained by
integrating  \equ{growth}.   As expected,  this  function is  strongly
dependent on the time  available for viscosity to redistribute angular
momentum.  At  high $z$  there is a  rapid growth, which  flattens off
toward   lower    redshifts.  The   horizontal    band   marks   the
observationally-inferred  density   of  mass  in   black  holes  today
\citep{S82, CT92,  KR95, FI99, SETAL99, MF01,YT02}. The  total mass in
black holes increases  with time, until the cool  gas supply stops due
to  some   feedback  mechanism  that  prevents   and/or  destroys  the
production of  molecular hydrogen.   Here we assume  that cosmological
reionization at  $\zre$ can be responsible for  preventing the further
cooling and  collapse of  haloes of mass  $\Mcrit$ for redshifts  $z <
\zre$.  This radiative feedback is expected to be especially effective
in  the relatively  small  systems under  consideration.   The $\pm  1
\sigma$ range  for the  redshift of reionization  as measured  by WMAP
($\zre =  17 \pm 5$, \citealt{S03})  is marked by  the vertical lines.
We expect the model prediction for  the seed mass density at $\zre$ to
be roughly  an order  of magnitude below  the accepted  observed value
today, in order to allow for the additional growth by accretion during
an AGN  phase.  We  see that for  $\zre=17$, say, this  requirement is
fulfilled once  viscosity operates  for a timescale  of $1\leq  t \leq
5$Myr.

Also  shown in  
Fig.~6
is the  case  $\xi =  1.3$  with a  viscous
timescale of $t = 20$ Myr.   The curve is flatter in this case because
$\Mcrit$ for  this profile  grows faster than  the $\xi =1$  case (see
figure 4 ), and as a result the contribution from new seed black holes
is smaller.  In  addition, the mass of the  black hole associated with
each halo above $\Mcrit$ scales as $\propto (1+z)^{3.8}$ for $\xi=1.3$
compared  with   $\propto  (1+z)^{1.5}$  for   $\xi=1$  (see  equation
\ref{eq:bhmass2}).  We  see that  the model predictions  for $\xi=1.3$
are acceptable  only for long  star-formation times, $t \geq  20$ Myr.
Flatter profiles,  with $\xi$ significantly larger  than $1.3$, either
require a very long viscous timescale for seed formation, or they fail
to produce enough mass in  black holes by the mechanism proposed here.
In this case one has only small early seeds and therefore must accrete
for many e-folding times.


\section{black holes and spheroids} 
\label{sec:correlation}

Our model  provides a prediction for  massive seed black  holes with a
characteristic linear correlation between the masses of the seed black
holes and  their host haloes.  In  this section we show  how our model
for high-redshift seeds  may lead naturally to a  relation between the
black hole mass and stellar  spheroid mass at $z=0$ by following their
evolution  through halo merger  trees.  We  associate a  galaxy's disc
mass  with gas that  is accreted  as diffuse  material since  the last
major merger event.   Each incoming halo is assumed to  host a mass of
cold gas in  its center, which can contribute  to a spheroid component
if  the  merger  occurs  on  a  short  dynamical  friction  timescale.
Similarly, if the merged halo hosts  a black hole seed (see below) and
reaches the central object in time,  then this seed can merge with the
central black  hole.  In  this fiducial example,  we assume  that when
black holes merge, their masses add linearly.

We  generate  halo  merger  trees  using the  algorithm  developed  by
\citet{SK99}  based on  the extended  Press-Schechter
formalism \citep{PS74, BCEK91, LC93}.  According to
this algorithm, the probability for a mass increase by $\Delta M$ in a
time step $\Delta t$ is
\begin{equation}
P(\Delta \sigma^2, \Delta  \delta_c) d (\Delta \sigma^2)= \frac{\Delta
\delta_c}{\sqrt{2  \pi}  \Delta \sigma^3}  e^{-\Delta  \delta_c^2 /  2
\Delta \sigma^2} \, d (\Delta \sigma^2),
\end{equation}
where $\Delta \delta_c = \delta_c(t)  - \delta_c(t + \Delta t)$ is the
linear fluctuation growth according  to the chosen cosmological model,
corresponding  to the  time step  $\Delta t$,  and 
$\Delta  \sigma^2 = \sigma^2(M) -  \sigma^2(M+\Delta M)$  
is the increment  in mean-square
fluctuation amplitude  according to  the given linear  power spectrum,
corresponding to  the mass increment  $\Delta M$.  We choose  the time
step such that
\begin{equation}
\Delta \delta_c \le \sqrt{ \Mmin \frac{d\sigma^2(M)}{dM}}
\end{equation}
for a chosen minimum halo mass $\Mmin$; a mass increase by $\Delta M <
\Mmin$ is  considered to be a  smooth accretion process  rather than a
merger  event.  The  merger tree  is generated  by  randomly selecting
progenitor masses  according to the above  probability distribution at
each  discrete time.   The parent  halo  is identified  with the  most
massive progenitor at  each time while the rest of  the mass is either
in  subhaloes  or diffuse  mass.   The tree  is  truncated  at a  high
redshift when the  mass of the major progenitor  is less than $\Mmin$.
We  run 50  Monte  Carlo  realizations of  the  merging histories  for
today's  haloes in  the mass  range  $10^{11} -  10^{14} \Msun$,  with
$\Mmin=10^6 \Msun$.

For each merger, we take into account the finite time for the incoming
subhalo to spiral in due to dynamical friction and reach the center of
the host  halo.  We  model each halo  as a singular  isothermal sphere
with a radius and velocity  given by the virial relations discussed in
\S 2.  The dynamical-friction time can then be written as \citep{BT87,
LC93},
\begin{equation}
t_{\rm df} = 0.42  \, {\rm Gyr} \, \left[ \frac{g(\epsilon,\eta)}{0.3}
        \right]  \, \left( \frac{R}{{\rm  ln} \Lambda}  \right) \left(
        \frac{1}{1+z} \right),
\end{equation}
where $R$ is defined as the  ratio of the mass of the major progenitor
to the mass of the  incoming halo.  For our calculation we approximate
the  Coulomb logarithm  as ${\rm  ln}\Lambda \simeq  {\rm  ln}R$.  The
function $g(\epsilon,\eta) =  \epsilon^{0.5} \eta^2$ characterizes the
effect   that  non-circular   orbits   have  on   the  orbital   decay
\citep[e.g.][]{T97,  GETAL98}.    Here  $\epsilon$  characterizes  the
circularity of the  orbit, and is defined as the  ratio of the angular
momentum of the orbit to the angular momentum of a circular orbit with
the same energy. The parameter  $\eta$ characterizes the energy of the
orbit by  defining $r_E  = \eta R_v$  to be  the radius of  a circular
orbit with  the same energy  as the actual  orbit.  We will  adopt the
typical  parameters   $\eta  =  0.6$   and  $\epsilon  =   0.5$  (see,
e.g., \citealt{ZB03}).

Material that ends up in a disc is assumed to be the gas that comes in
as  diffuse accretion  since the  last major  merger.  The  last major
merger is defined as a merger with an incoming object that is at least
$50 \%$  the mass  of the  host. 
The diffuse  gas may partly be  the gas lost from  dark haloes smaller
than $M \sim 2 \times 10^8 [(1+z)/18]^{-3/2}$ due to photo-evaporation
starting   at  the   epoch  of   reionization   (e.g.   \citealt{B00,
Shaviv04}).  Although  the gas escapes  from the small haloes,  it may
remain  bound to  the  larger host  halo,  and we  assume  that it  is
accreted onto the disc at about  the same time the low-mass haloes are
accreted.  For  haloes less massive than  $\sim 3\times 10^{11}\Msun$,
which roughly  coincides with  the upper limit  for disc  formation at
high redshift,  we indeed expect  short infall times because  of short
cooling times and negligible shock heating \citep{BD03}.

The spheroid mass is assumed to be the straightforward sum of the cold
baryons  in the merged  subhaloes throughout  the merger  history that
sank to the  center of the subhalo.  We  take the spheroidal component
to be in place by the  redshift of the last major merger assuming that
a major merger would disrupt  the system into an irregular galaxy.  If
the estimated sinking time of the incoming subhalo is shorter than the
time interval between the subhalo  merger and the last major merger we
add to  the bulge a  mass $f_0 \Ms$,  where $f_0=0.03$ is  the assumed
cold-gas  fraction, and $\Ms$  is the  mass of  the merged  halo.  The
choice  of the mass  fraction in  the definition  of the  major merger
makes  only a  small difference  to the  final slope  of  the relation
between black hole and spheroid mass, because the merger history tends
to converge to the final mass after the last major merger.

Among the subhaloes  that reach the center before  the redshift of the
last major merger, we determine if  they have a black hole, and if so,
we  assume  it  merges  with  the  central black  hole  over  a  short
timescale.   In  principle,  we  could  compute the  black  hole  mass
contained in each  merged halo by following its  full merger tree back
in time, until some early redshift $z \gsim \zre$.  However, since our
aim is  simply to  investigate the ramifications  of our  scenario, we
have  chosen  only to  follow  the merger  tree  of  the most  massive
progenitor, and estimate the formation times of each merged subunit in
a  statistical fashion.   This choice  saves a  significant  amount of
computational time.

Specifically, since  seed black holes stopped forming  at $z=\zre$, we
can estimate  the total black hole  mass present in a  subhalo of mass
$\Ms$ that  merges into  the main  host at a  redshift $\zmer$  by the
following approximation.  If $\zmer  < \zre$, then the black-hole mass
populating the  merging subhalo is determined by  integrating over the
mass function of its progenitors at $\zre$,
\begin{eqnarray}
\Mbh &=&  \int_{{\Mcrit(\zre)}}^\infty \frac{\Ms}{M} \, \Mbh(M,\zre)
\nonumber  \\  &\times&   f_2  [  \sigma(M),  \zre  \,|\,
\sigma(\Ms),  \zmer ]  \nonumber  \\ 
&\times&  | \frac{  d \sigma^2(M)}{d M} | \, d M .
\end{eqnarray}
Here, $M$ is  the mass of progenitors.  The  function $f_2 [\sigma(M),
\zre \,|\,  \sigma(\Ms), \zmer ]$  represents the fraction of  mass in
haloes of  mass $\Ms$  at $\zmer$ that  was in  haloes of mass  $M$ at
redshift  $\zre$  (see  equation  2.15  of  \citet{LC93}).   The  mass
$\Mbh(M,\zre)$ is the  mass of the seed black hole  residing in a halo
of  mass $M$  at $\zre$  according  to \equ{bhmass2}.   If the  merger
occurs  before $\zre$,  then the  black-hole residing  in  the merging
subhalo is  assumed to have  a mass according to  \equ{bhmass2}, given
the halo  mass $\Ms$ at  $\zmer$.  We only  need to follow  the merger
history back  to the  time when  the major progenitor  is as  small as
$\Mcrit(z)$, because  when a progenitor  is smaller than  $\Mcrit$, no
seed black holes will form.

The  integral above  gives  only an  average  black hole  mass for  an
ensemble of host haloes of  mass $\Ms$.  So the approximation is valid
if  $\Mbh \gg  M_0$,  where $M_0$  is  the seed  mass associated  with
critical-mass haloes at $\zre$.   However, for low-mass subhaloes, the
approximation can  break down because  it allows solutions  with black
hole masses smaller than the initial seed masses, $\Mbh < M_0$ ($ \sim
6 \times  10^4 \Msun$, see Eq.   18).  This is  unphysical, because we
assume that  each black hole is built  up by a series  of mergers with
seeds of mass  $M_0$.  We attempt to remedy  this shortcoming in cases
by  randomly  re-assigning  the   value  of  $\Mbh$  in  these  cases.
Therefore if  $\Mbh < M_0$ we  choose a random uniform  deviate r, and
set $\Mbh = 0$ if $r > \Mbh/M_0$ and $\Mbh = M_0$ otherwise.

\begin{figure}
\includegraphics[width=84mm]{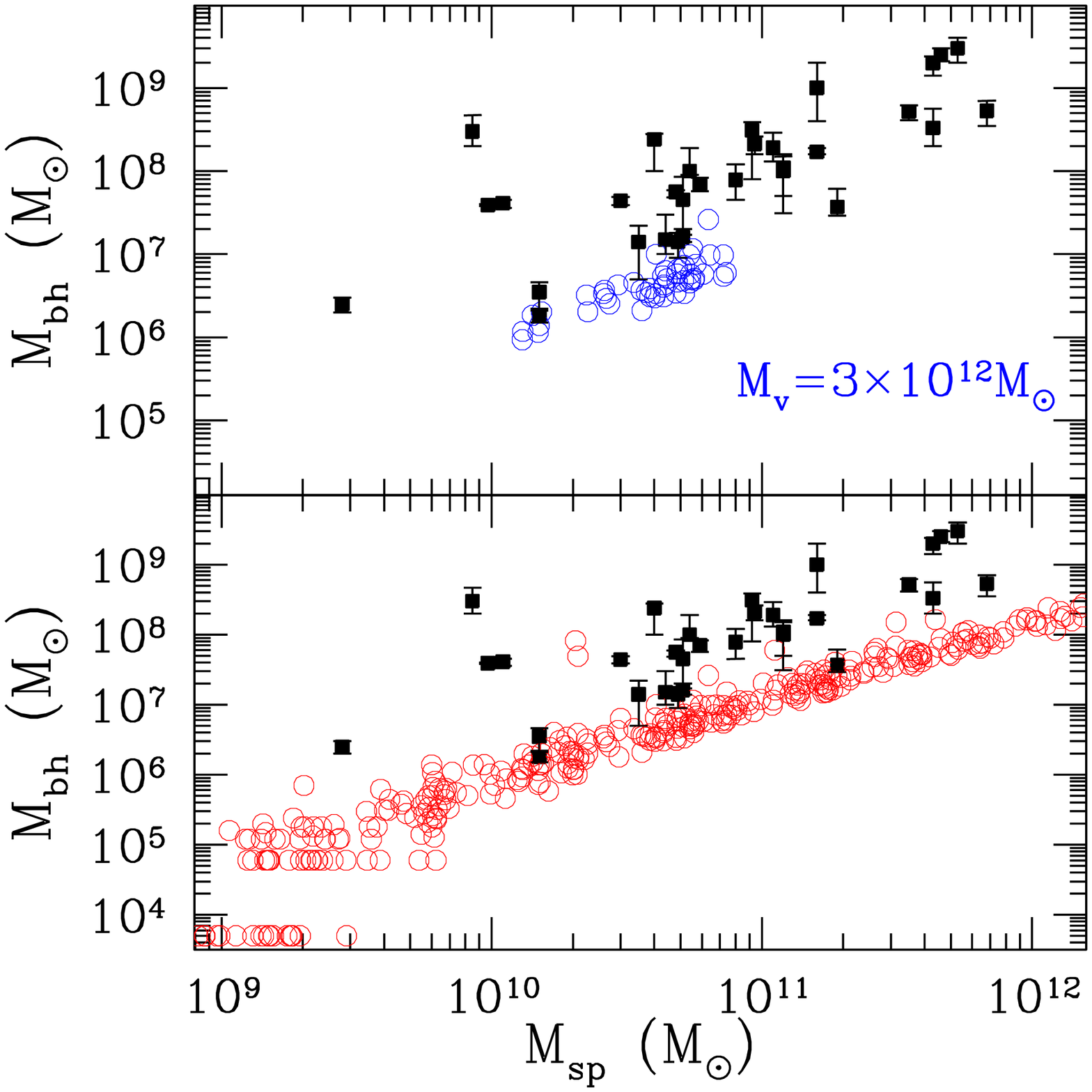}
\caption{  Today's  black-hole mass  versus  host-spheroid mass.   The
filled symbols are based on data from \citet{TETAL02}, where the bulge
mass is  derived from the  B absolute magnitude assuming  the constant
mass-to-light ratio advocated by \citet{VDM91}.  The model predictions
are based on merger growth only, ignoring accretion.  Top: for a fixed
halo mass  today, $\Mv  = 3.2 \times  10^{12} \Msun$, using  50 random
realizations of the merger tree.   The variations in merger history by
themselves  tend to  spread the  masses along  the  observed relation.
Bottom: for a  range of present halo masses,  $\Mv = 10^{11}-10^{13.5}
\Msun$, with logarithmic half order  of magnitude steps.  The slope of
he predicted  relation is  in agreement with  the observed  slope.  In
cases  where the black-hole  mass is  calculated to  be zero,  we have
placed  the  symbol at  the  bottom of  the  panel  for clarity.   The
predicted  black-hole masses  could potentially  increase by  about an
order of magnitude due to accretion in an AGN phase.}
\label{fig:fig7}
\end{figure}

Figure 7
shows  the  resulting black-hole  mass  as  a
function of the  host spheroid mass from the  merger tree realizations
described above  (open symbols), in comparison with  the observed data
(solid  points  with errors,  see  caption).   We  have adopted  model
parameters of $t=10$ Myr and  $\zre=15$.  The model predictions in the
top  panel  refers  to  a  given  {\textit{fixed}}  halo  mass  today,
$M=3.2\times 10^{12}\Msun$,  where only  the random merger  history is
varied from  point to  point.  Though not  shown, the  associated disc
masses for these  haloes range from $10^{10} -  10^{11} \Msun$ with no
clear correlation between the disk and black hole masses.

We  find  that  the black  hole  mass
correlates with the mass of the spheroid  component, as observed, even
for a fixed-mass host halo.  This result arises because subhaloes that
are massive enough  to  sink to  the  center of  the parent  halo and
contribute to a bulge,  are
generally massive enough  to host black holes  as  well.  Haloes which
experienced  many mergers  with small subhaloes  tend  to have a small
spheroids  and small black   holes.
This would naturally  explain why
black hole mass is  observed to correlate  with spheroid mass and  not
disc mass\citep{GETAL01}, since disc mass does not correlate well with
the merger history in our picture.  

The  bottom panel  of  
Fig.~7
shows several  ensemble
realizations for  a series  of host halo  masses with $z=0$  values of
$10^{11}  - 10^{13.5}  \Msun$. The  computed relation  follows closely
$\Mbh  \propto \Msp$,  and is  qualitatively similar  to  the observed
relation at $z=0$.   The normalization (set primarily in  this case by
setting $t=10$Myr)  allows for an  additional growth by a  few e-folds
during  an AGN  accretion phase,  as  desired.  In  addition, for  the
lowest-mass spheroids, we  see in some cases the  absence of any black
hole mass (points  along the bottom edge represent  $\Mbh = 0$).  This
comes about  because low mass host haloes  are less likely to have had
progenitors  more  massive than  $\Mcrit$  at  $z=\zre$.  Black  holes
cannot be  smaller than  the characteristic seed  mass $\sim  6 \times
10^4 \Msun$ while there is no fixed mass threshold for a progenitor to
contain gas  that will  eventually end up  in a spheroid.   This gives
rise to a  slightly steeper slope in the bottom  left of the relation.
The  apparent increase in  the number  of low  mass spheroids  with no
black  holes also  suggests  the presence  of  a lower  mass limit  on
spheroids  hosting black  holes, of  roughly $\Msp  \sim  10^8 \Msun$.
This is a potentially verifiable prediction of the model.

It  is  of interest  to  investigate  the  robustness of  the  derived
relation  between $\Mbh$  and $\Msp$  to specific  assumptions  of the
model  presented  here.   For   example,  the  result  is  practically
independent  of the  finite  duration  of a  merger  due to  dynamical
friction.  Major mergers are quick and have an immediate effect on the
growth rate  of the spheroid  as well as  the black hole,  while minor
mergers  take a long  time to  complete and  therefore have  a delayed
effect that  may not materialize  in a Hubble time  \citep{ZB03, TB01,
TB03}.  When we repeat the  calculation under the assumption that {\it
all} mergers  are instantaneous, ignoring  the finite duration  of the
spiral-in process,  we find that the predicted  relation between black
holes and spheroids remains almost unchanged, with only a slight shift
of typically  less than  10\% towards larger  spheroid masses  for the
same black hole  mass.  This 
small shift 
is to be expected  since the less massive
subhaloes that  now contribute to the  spheroid mass tend  to be below
the critical mass for seed formation  and thus do not add to the black
hole mass.

We find the  predicted slope of the relationship  to be insensitive to
the  actual value  of $\Mcrit$  (as long  as $\Mv  \gg  \Mcrit$).  The
normalization of the relation, namely  the black-hole mass for a given
spheroid mass,  is roughly inversely proportional  to $\Mcrit$.  This
is because for a lower value  of $\Mcrit$ more haloes are able to form
a seed and contribute to  the merged black-hole mass without affecting
the spheroid mass.

We also find that the slope  is insensitive to the value of $\zre$ (as
long as  $\zre \geq  6$), while the  normalization does depend  on it.
For a lower $\zre$, seed formation takes place in a larger fraction of
the haloes that eventually merge to form spheroids.  By lowering $\zre$
from 15  to 6 we obtain  an order of magnitude  increase in black-hole
mass.  A  comparable decrease  in  black-hole  mass  is obtained  when
increasing $\zre$ from 15 to 20.

However, the slope of the relation  is found to be mildly sensitive to
the original  scaling between  the seed black  hole mass and  its host
halo  mass.  If  we replace  the  predicted correlation  of our  model
($\Mbh \propto \Mv$)  with an ad hoc seed  assignment to critical-mass
haloes at  high redshift, e.g.,  $\Mbh \propto {(\Mcrit)}^3$,  we find
that the linear $\Mbh \propto \Msp$ relation steepens to roughly $\Mbh
\propto  \Msp^{3/2}$.  This  is because  the ratio  $\Mbh/\Mv$  is now
monotonically increasing with $\Mv$.  This mild sensitivity means that
the  predicted  correlation  of   our  physical  model  for  seeds  is
important, but it does not have to be exactly $\Mbh \propto \Mv$.

Finally, we point out that our result is quite sensitive to the way we
assign mass $M$  to a black hole which results from  the merger of two
black holes, of  masses $M_1$ and $M_2$.  In  our fiducial calculation
we  have  assumed  that  the   black  hole  masses  sum  up  linearly,
$M=M_1+M_2$.  On the other hand,  if entropy is conserved, the mass of
the merger product  is $M = \sqrt{M_1^2 +  M_2^2}$ \citep{H71a, H71b}.
This should serve as a lower  bound to the possible mass of the merger
product,  and  should  yield   the  flattest-possible  $\Mbh  -  \Msp$
relation.  Consider a black hole in the center of the halo formed from
$N_{\rm  m}$ mergers  of  seed black  holes  of mass  $\Mbh \sim  10^5
\Msun$.  If entropy is conserved, and $N_{\rm m} \gg 1$ then the black
hole mass would be $\Mbh =  M_{\rm seed} (1 + N_{\rm m})^{1/2} \propto
N_{\rm  m}^{1/2}$.  In  our  model,  spheroids grow  via  a series  of
mergers with the  gas from nearly equal-mass haloes,  roughly as $\Msp
\simeq  f_0 M_{s} N_{\rm  m} \propto  N_{\rm m}$,  so we  expect $\Mbh
\propto \Msp  ^{1/2}$ in this limit.   When we follow  this recipe for
mass  adding using  our full  merger-tree method,  we indeed  obtain a
result  very close  to  our expectation:  $\Mbh \propto  \Msp^{0.55}$.
Thus, in  this limit, the  derived relation would be  somewhat flatter
than observed.   In order  to comply with  the observed  relation, one
would  have  to appeal  to  a  scenario  where black-hole  growth  via
accretion  is  more  efficient  in  high-mass  spheroids  \citep[e.g.,
][]{VHM03}.

With  the above tests  in mind,  we conclude  that the  derived linear
relation between black hole and spheroid mass depends primarily on the
fact that the  seeds were massive, that they  formed at high redshift,
and that their  masses add roughly linearly when  they merge.  In this
case the build-up of black holes and spheroids is occurring in concert
through  mergers, thus  setting  the linear  relationship.  A  similar
linear relation  would hold for  the final black-hole  masses provided
that it has  not been changed by the  subsequent growth experienced by
black holes  via accretion.  If  the accretion is proportional  to the
mass of the seed, as would  be the case for accretion at the Eddington
rate, then we expect the slope of the relation to persist.  Of course,
the effect  of accretion  depends on what  fraction of the  black hole
mass was accreted.  If the increase in  mass is only by a  factor of a
few, then  one could assume  that the role  of mergers in  setting the
observed relationship is  important. On the other hand,  if the growth
by accretion is by more than an order of magnitude, then the accretion
process would have  the dominant role in determining  the slope of the
relation.  According  to our model, the normalization  of the observed
relationship  is determined  additively  by two  free parameters:  the
viscous  timescale and  the  growth via  accretion.   When adopting  a
physically  motivated value for  the viscous  timescale, we  find that
growth via  accretion is  limited to  be by an  order of  magnitude or
less,  thus suggesting  that  mergers  do play  an  important role  in
setting the relationship.  We note that in models  where the seeds are
of  smaller masses,  $\sim 100\,  \Msun$ (e.g.,  \citealt{VHM03}), the
growth by  mergers is negligible and  the accretion process  has to be
tuned with a specific feedback scenario in order to obtain the correct
relation between black holes and spheroids.

Although we focused here on black  hole and spheroid growth, we do not
expect all the  seed black holes accreted by a dark  matter halo to be
incorporated  in  the spheroid.  A  number  of  black holes  could  be
orbiting  in haloes  of  big  galaxies. They  could  have arisen  from
subhaloes that  were disrupted after  entering the host halo.   Such a
scenario  has been  investigated  by \citet{ITS03a}  and the  possible
observational consequences are explored in \citet{ITS03b, ITS03c}.

%
%

\section{Conclusion} 
\label{sec:conc}
 
We presented a physical model for the production of massive seed black
holes  at  high  redshifts,  with  a characteristic  mass  $\sim  10^5
\Msun$. We argued that this model provides a useful starting point for
explaining the  observed relation between the masses  of today's black
holes and  the spheroidal components  of galaxies.  Massive  seeds may
also  help  explain the  existence  of  the  supermassive black  holes
associated with luminous AGN at $z \gsim 6$.

Our model  for seed formation  is based on  the general idea  that the
mass   in  every  early-collapsing   halo  is   expected  to   have  a
{\textit{distribution}}  of specific angular  momentum, with  at least
some  fraction of  its  gas  having a  much  smaller specific  angular
momentum than  is characterized  by the global  spin parameter  of the
halo.   After the low-spin  gas cools,  it can  fall into  the central
region and form  a dense disc, where certain  viscous processes become
effective and may potentially lead to black hole formation.

In our exploration of this  idea, we assumed that the specific angular
momentum of the gas is similar to that measured for the dark-matter in
N-body simulations.   In this case,  if angular momentum  is conserved
during the  gas infall, the central gas  discs become self-gravitating
and Toomre  unstable in  all the haloes  above a critical  mass.  This
critical mass  has a median value  of $\sim 7 \times  10^7\Msun$ at $z
\sim 15$,  and it is  expected to be  scattered log-normally with  a ln
standard deviation  of $\sim 0.9$.   The sites of seed  production are
thus much more massive than  the typical collapsing mass at that time,
and are therefore associated  with high-sigma peaks in the fluctuation
distribution.   Gravitational  instabilities  in  the  disc  drive  an
effective viscosity that redistributes angular momentum.  As a result,
material from the low angular  momentum tail of the distribution loses
its  angular momentum  on a  viscous timescale  and contribute  to the
formation of a  black hole.  We associate this  viscous timescale with
the  timescale  for disruption  and/or  heating  of  the disc  due  to
substantial star  formation.  We  expect this timescale  to be  on the
order of $1-30$ Myr, but its exact value is not well determined.

We found  that the  typical seed black  holes, which form  in critical
mass haloes,  have a characteristic  mass of $\sim 10^5  \Msun$, quite
insensitive to the epoch of formation.  This predicts a lower bound to
possible masses of  massive black holes, which can  be confronted with
observations today.

This scenario should operate continuously as additional haloes grow to
above the critical mass and  create new potential sites for black-hole
formation.   The  process  is  expected  to stop,  however,  when  the
Universe becomes sufficiently reionized that ${\rm H}_2$ molecules are
destroyed and  cannot provide efficient cooling.   We therefore expect
seed black holes to form before $\zre \sim 13-21$.

We showed that the comoving density  of mass in seed black holes grows
gradually  with time, in  association with  the birthrate  of critical
mass  haloes.  This density  becomes comparable  to the  observed mass
density of supermassive black holes  by $z \sim 15$, near the redshift
of reionization.  Depending on the adopted viscous timescale, there is
room for an order of magnitude  growth in black hole masses via quasar
powering accretion.  We find that  a viscous timescale of $t\sim 5-10$
Myr provides quite reasonable results.

We explored  the development of a correlation  between black-hole mass
and spheroid  mass in today's  galaxies by tracking our  early forming
massive  seed  population  through  a  halo merger  tree.   Under  the
assumption that spheroids  form via major mergers, we  showed that the
merger process leads naturally to a linear relation between black-hole
mass and spheroid mass.  This result is mainly due to the formation of
{\textit{massive}}  seeds  only  in  rare, high-peak  haloes  at  high
redshifts,  and  is not  too  sensitive to  the  details  of our  seed
formation model.  Starting with $\sim 1-100\msun$ seeds would not lead
to  a  linear  correlation  via  mergers  alone;  additional  feedback
associated with AGN accretion is  needed.  The seed black holes in our
model have masses that are  proportional to the virial masses of their
host haloes  at formation.  We demonstrated that  this linear relation
is preserved  if black  holes merge in  concert with  the hierarchical
growth of haloes and spheroids.

One  potentially interesting ramification  of our  scenario is  that a
population  of massive  black holes  at high  redshift might  serve to
power  small  quasars,  fueled  perhaps  by  low-angular-momentum  gas
liberated by  merger events that occur  after the seeds  are in place.
\citet{MAD03} discussed how a population of ``mini-quasars'' like this
might be  responsible for  reionizing the universe  at $\zre  \sim 15$.
However,   in  their   model,  the   mini-quasar   power-sources  were
intermediate-mass $\sim 100 \Msun$ black holes.
Under similar assumptions,  the massive seed black holes  of our model
would produce  even more ionizing  photons per site, and  perhaps lead
more naturally to an early  reionization.  In our current scenario the
reionization epoch was an input parameter detached from the black-hole
formation  process, but, in  principle, the  accretion onto  the seeds
themselves could  be responsible for the  reionization that eventually
inhibits further black-hole formation.

Our quantitative  predictions should not be  interpreted too literally.
For   example,   our  fiducial   model   assumes   that  the   initial
angular-momentum distribution of the gas mirrors that of a typical CDM
halo, and  that it  is preserved during  disk formation.  There  is no
evidence in local  galaxies for discs with inner  densities as high as
invoked by our  model for proto-galaxies.  This might  be explained by
processes  that  have altered  the  gas angular-momentum  distribution
preferentially  at  late  epochs,   such  as  feedback  effects  (e.g.
\citealt{MD02}), but one should clearly assign a substantial degree of
uncertainty  to  our  model  assumption.   In order  to  address  this
uncertainty, at least qualitatively, we explored the effect of varying
the initial angular  momentum distribution.  We found that  if the low
angular  momentum tail  is as  minor as  in an  exponential disc  of a
constant-density core  ($\xi =2$), the resulting seed  black holes are
too small to be of relevance.  An intermediate case, with $\xi = 1.3$,
can produce sensible black-hole seeds, comparable to our fiducial case
($\xi  =  1$), provided  that  the  viscosity  timescale is  slightly
longer, $20 \le t \le 30$  Myr.  For $\xi=1.3$, the critical halo mass
becomes $\sim 10^8 \Msun$ at $z=15$,  and the inner $\sim 10\%$ of the
disc  mass manages  to  lose  its angular  momentum  by viscosity  and
produce a  black hole of  mass $\sim 10^6  \Msun$.  This seed  mass is
larger than in the fiducial case, but the seeds in this case are rarer
events (see figure \ref{fig:fig4}).   Therefore, in order to recover a
similar total mass density in  black holes, the viscosity must operate
for a  longer duration.  The  actual low angular-momentum tail  of the
gas  distribution   remains  an  unknown,  which   will  hopefully  be
constrained  by  future,  more realistic  hydrodynamical  simulations,
properly including feedback and other important physical effects.

In conclusion, while the model presented here is clearly idealized, it
does  provide a useful  example for  the kind  of mechanism  needed to
connect  cosmological  structure   formation  on  galactic  scales  to
black-hole formation on much smaller scales.  Observations suggest the
presence of  massive black holes  at high redshifts and  therefore the
need for  the formation of massive  seed black holes  at still earlier
times.  According to our  scenario, the observed correlations of black
hole  and galaxy properties  are directly  linked to  the hierarchical
nature of structure formation itself,  and owe their existence to rare
systems forming in association with the first light.

%
%
\section{Acknowledgments} 
 
We acknowledge  stimulating comments from Martin  Haehnelt,  
Paul Martini,  Eve  Ostriker, Martin  Rees, Brant  Robertson,
David Weinberg, Terry Walker and  Andrew Zentner.  
We thank  the anonymous referee for helpful  suggestions that improved
the quality  of this article.   SMK thanks the 2003 Santa  Fe cosmology
workshop where  part of  this work was  completed. AD thanks  the IAP,
Paris, for their hospitality.  This work has been supported by U.S.DOE
Contract  No.  DE-FG02-91ER40690 (SMK),  NASA Hubble  Fellowship grant
HF-01146.01-A (JSB), and ISF grant 213/02 (AD).

%
%

\label{lastpage}
                                    
\end{document}